\documentclass[conference, 10pt]{IEEEtran}

\newcommand{\ignore}[1]{}
\usepackage{fancyhdr}
\usepackage[normalem]{ulem}
\usepackage[hyphens]{url}

\usepackage[top=0.75in, bottom=1in, left=0.625in, right=0.625in]{geometry}

\usepackage[normalem]{ulem}

\usepackage{epsfig}
\usepackage{amssymb} 
\usepackage{booktabs}
\usepackage{amsmath} 
\usepackage{amsthm}
\usepackage{soul}

\usepackage{graphics,eepic,epic}
\usepackage{latexsym}
\usepackage{euscript}
\usepackage{tabularx}
\usepackage{graphics,eepic,epic,psfrag}
\usepackage{color}
\usepackage{algorithm, algorithmic}
\usepackage{tabularx}
\usepackage{multirow}
\usepackage{xspace}
\usepackage{times}
\usepackage{color}
\newcommand{\yanpei}[1]{{\color{blue} [#1 -- Edited by Yanpei]}}

\newcommand{\powerscale}{{FastCap}\xspace}
\newtheorem{theorem}{Theorem}

\usepackage{epstopdf}

\begin{document}

%%%%%%%%%%%---SETME-----%%%%%%%%%%%%%
\title{FastCap: An Efficient and Fair Algorithm for Power Capping in Many-Core Systems}
%%%%%%%%%%%%%%%%%%%%%%%%%%%%%%%%%%%%

\author{
\IEEEauthorblockN{Yanpei Liu\IEEEauthorrefmark{1},
Guilherme Cox\IEEEauthorrefmark{2},
Qingyuan Deng\IEEEauthorrefmark{3}, 
Stark C. Draper\IEEEauthorrefmark{4} and
Ricardo Bianchini\IEEEauthorrefmark{5}}
\IEEEauthorblockA{\IEEEauthorrefmark{1} Facebook Inc. and University of Wisconsin Madison,
{\em yanpeiliu@fb.com}}
\IEEEauthorblockA{\IEEEauthorrefmark{2} Rutgers University, {\em guilherme.cox@rutgers.edu} }
\IEEEauthorblockA{\IEEEauthorrefmark{3} Facebook Inc., {\em qdeng@fb.com}}
\IEEEauthorblockA{\IEEEauthorrefmark{4} University of Toronto, {\em stark.draper@utoronto.edu}}
\IEEEauthorblockA{\IEEEauthorrefmark{5} Microsoft Research, {\em ricardob@microsoft.com}}
}

\maketitle

\begin{abstract}

Future servers will incorporate many active low-power modes for
different system components, such as cores and memory. Though these
modes provide flexibility for power management via Dynamic Voltage and
Frequency Scaling (DVFS), they must be operated in a coordinated
manner. Such coordinated control creates a combinatorial space of
possible power mode configurations. Given the rapid growth of the
number of cores, it is becoming increasingly challenging to quickly
select the configuration that maximizes the performance under a given
power budget.  Prior power capping techniques do not scale well to
large numbers of cores, and none of those works has considered memory
DVFS.

In this paper, we present \powerscale, our optimization approach for
system-wide power capping, using both CPU and memory DVFS.  Based on a
queuing model, \powerscale formulates power capping as a
non-linear optimization problem where we seek to maximize
the system performance under a power budget, while promoting fairness
across applications.  Our \powerscale algorithm solves the
optimization online and efficiently (low complexity on the number of
cores), using a small set of performance counters as input.  To
evaluate \powerscale, we simulate it for a many-core server running
different types of workloads.  Our results show that \powerscale caps
power draw accurately, while producing better application performance
and fairness than many existing CPU power capping methods (even after
they are extended to use of memory DVFS as well).
% \powerscale also has a very low algorithmic complexity, suggesting a
% promising direction for future power management in many-core systems.

\end{abstract}

%\category{CR-number}{subcategory}{third-level}

% general terms are not compulsory anymore, 
% you may leave them out
%\terms
%term1, term2

%\keywords
%keyword1, keyword2

% \category{D.4}{Operating systems}{Storage management}
% \terms{Design, experimentation}
% \keywords{Energy management, energy modeling, disk energy}

\section{Introduction}

% RB: Reducing the focus on datacenter to avoid complaints about
% not showing results for services and other more modern applications.
% Datacenters consume massive amounts of power.  According to
% J. Koomey's estimates for 2010 \cite{Koomey11}, datacenters consumed
% roughly 1.5\% and 2\% of the total electricity consumed world-wide and
% in the United States, respectively.  To reduce their operational cost
% and impact on the environment, it is critical to manage their power
% consumption \cite{Barroso07, Chase01, Chen05, Heath05, Lefurgy03,
%  Meisner11, Pinheiro01}.  As practical evidence of this need, 
As power and energy become increasingly significant concerns for
server systems, servers have started to incorporate an increasing
number of idle low-power states (such as CPU sleep states) and active
low-power modes of execution (such as CPU DVFS states). 
Researchers have also proposed active low-power modes
for the main memory subsystem \cite{David11, Deng12a, Deng11}, for
disk drives \cite{Carrera03, Gurumurthi03}, and for interconnects
\cite{Abts10}.  Liu {\em et al.}~\cite{SleepScale, LiuCISS}
and Vega {\em et al.}~\cite{Vega13} showed that CPU active low-power
modes and idle low-power states can jointly achieve high energy
efficiency.  In contrast, Meisner {\em et al.}~suggested that active
low-power modes are the only acceptable alternative for conserving
energy in the face of interactive workloads \cite{Meisner11}.  Deng
{\em et al.}~showed that the management of CPU and memory active
low-power modes must be coordinated for stability and increased energy
savings \cite{Deng12b}.  Since the CPU power modes affect the traffic
seen by the memory subsystem and the memory power mode affects how
fast cache misses are serviced, a lack of coordination may leave the
system unable to properly manage energy consumption and performance.

\iffalse \yanpei{To see an example of such behavior, consider
  Figure~\ref{fig::algo}. Suppose the overall goal is to maximize the
  full-system energy savings within a pre-selected application
  performance degradation bound.}  \fi

Like \cite{Deng12b}, we consider both CPU and memory active
low-power modes.  However, instead of maximizing energy savings within
a performance bound, {\em we consider maximizing application
performance under a full-system power consumption cap/budget.}  Such
power capping is important because provisioning for peak power usage
can be expensive, so designers often want to oversubscribe the power
supply at multiple levels \cite{Barroso13,Wang13}.

A lack of coordination hampers a system's ability to maximize
performance under a full-system power cap.  To see an example, suppose
that the applications are mostly memory-bound, and just changed
behavior, causing the system power consumption to decrease
substantially below the power budget.  In this situation, the CPU
power manager (which does not understand memory power and assumes that
it will stay the same regardless of the cores' frequencies) might
decide that it could improve performance by increasing the core
voltage/frequency and bringing the system power very close to the
budget.  The near-budget power consumption would prevent the
(independent) memory power manager from increasing the memory
frequency.  Adhering to the power budget in this way would produce
more performance degradation than necessary, since the applications
would have benefited more from a memory frequency increase than core
frequency increase(s).  The situation would have been better, if the
memory power manager had run before the CPU power manager.  However,
in this case, a similar problem would have occurred for CPU-bound
applications.
% \ricardo{I'm not entirely happy with this example.}

% These behaviors suggest that it is essential to coordinate
% power-performance management across system components to ensure that
% the system is balanced and yields the desired behavior.  
Coordination is especially important when maximizing performance under
a server power cap for three reasons: (1) exceeding the server power
budget for too long may cause temperatures to rise or circuit breakers
to trip; (2) it may be necessary to purchase more expensive cooling or
power supply infrastructures to achieve the desired application
performance; and (3) even when the power capping decisions are made at
a coarser grain (e.g., rack-wise), individual servers must respect
their assigned power budgets.

\iffalse
\begin{figure}
\centering
\includegraphics[width = 0.5\textwidth]{figures/example.eps}
\caption{Comparison between different core/memory frequency coordination 
schemes. Semi-coordinated approach causes performance to fluctuate, 
while fully coordinated minimizes performance fluctuations.}
\label{fig::algo}
\end{figure}
\fi

The abundance of active low-power modes provides great flexibility in
performance-aware power management via DVFS. However, {\em the need
for coordinated management creates a combinatorial space of possible
power mode configurations.}  This problem is especially acute for
future {\em many-core servers,} especially when they run many
applications (each with a potentially different behavior), since it is
unlikely that the power mode selected for a core running one
application can be used for a core running another one.  Quickly
traversing the large space of mode combinations to select a good
configuration as applications change behavior is difficult.  For small
core counts and in the absence of memory DVFS, Isci {\em et al.}
\cite{Isci06} proposed exhaustive search for the challenging scenario
in which the server runs as many applications as cores.  The time
complexity of the search increases exponentially in the number of
cores and, thus, their approach does not scale to large core counts.
More recent works (e.g., \cite{Bose12,Sharkey07}) have improved on
Isci's exhaustive search, {\em but never addressed the combination of
CPU and memory DVFS.}  Moreover, most prior works attempt to maximize
instruction throughput, {\em which causes an unfair power allocation
across applications} (CPU-bound applications tend to get a larger share
of the power).

With these observations in mind, in this paper we propose \powerscale,
a methodology and search algorithm for performance-aware full-system
power capping via both CPU and memory DVFS. \powerscale efficiently
selects vol\-tage/frequency configurations that maximize a many-core
system's performance, while respecting a user-provided power budget.
Importantly, \powerscale also enforces fairness across applications,
so its performance maximization is intended to benefit all
applications equally instead of seeking only the highest possible
instruction throughput.  \powerscale has very low time complexity
(linear in the number of cores), despite the combinatorial number of
possible power mode configurations.

To devise \powerscale, we first develop a queuing model that
effectively captures the workload dynamics in a many-core system
(Section~\ref{sec.queuing_model}).  Based on the queuing model, we
formulate a non-linear optimization framework for maximizing the performance
under a given power budget (Section~\ref{sec.optimization}).  To solve
the optimization problem, we make a key observation that core
frequencies can be determined optimally in linear time for a given
memory frequency.  We develop the \powerscale algorithm
(Algorithm~\ref{alg.alg1}) and implement it to operate online. The
operating system runs the algorithm periodically (once per time
quantum, by default), and feeds a few performance counters as inputs
to it (Section~\ref{sec.implementation}).

We highlight two aspects of \powerscale: (1) it
does OS-based full-system power capping, as the performance- and
fairness-aware joint selection of CPU and memory DVFS modes is too
complex for hardware to do; and (2) it enforces caps at a relatively
fine per-quantum (e.g., several milliseconds) grain, as rapid control
may be required depending on the part of the power supply
infrastructure (e.g., server power supply, blade chassis power
supplies, power delivery unit, circuit breaker) that has been
oversubscribed and its time constants.  Moreover, capping power
efficiently at a fine granularity is more challenging than doing so at
a coarse one.  Nevertheless, \powerscale assumes that the server
hardware is responsible for countering power spikes at even shorter
granularities, if this is necessary.

To evaluate \powerscale, we simulate it for a server running different
types of workloads (Section~\ref{sec:evaluation}).  (A real
implementation is not possible mainly as \powerscale applies memory
DVFS, which has recently been proposed in \cite{David11,Deng11} and is
not yet readily available in commercial servers.) Our results show
that \powerscale maintains the overall system power under the budget
while maximizing the performance of each application.  Our results
also show that \powerscale produces better application performance and
fairness than many state-of-the-art policies (even after they are
extended to use memory DVFS as well), because of its ability to fairly
allocate the power budget and avoid performance outliers.  Finally,
our results demonstrate that \powerscale behaves well in many
scenarios, including different processor architectures (in-order
vs.~out-of-order execution), memory architectures (single vs.~multiple
memory controllers), numbers of cores, and power budgets.

%In
%fact, we discuss how to broaden the scope of FastScale for managing
%more resources, such as server disks and network interfaces, as they
%start to incorporate active low-power modes.

% RB: The main sections have already been listed above.
% The remainder of the paper is organized as follows.  The next section
% discusses the background and the related prior works, including
% CoScale.  Section \ref{sec:fastscale} describes the FastScale
% methodology and search algorithm.  Section \ref{sec:evaluation}
% presents our evaluation methodology and results.  Finally, Section
% \ref{sec:conclusion} draws our conclusions.

% {\bf RB: We need to revisit the memory timings we will use.  DDR3 at
% 800MHz, as used in the CoScale paper, is slow by today's standards.
% We should probably use 1066MHz, if we can find all the needed info
% about these parts.  Guilherme, can you please look into this?}

\section{Related Work and Contributions}

Though they have not considered memory DVFS, many prior works have
proposed using a global controller to coordinate cores' DVFS subject
to a CPU-wide power budget, e.g. \cite{Isci06, Bose12, Sharkey07}. Next, we
overview some of the works in this area.

\noindent{\bf Optimization approaches.}  Sharkey {\em et al.~}\cite{Sharkey07}
studied different designs and suggested that global power management
is better than a distributed method in which each core manages its own
power. They also argued that all cores receiving equal share of the
total power budget is preferred over a dynamic power redistribution,
due to the complexity of the latter approach.  Isci {\em et
al.~}\cite{Isci06} used exhaustive search over pre-computed power and
performance information about all possible power mode combinations.
Their algorithm's time and storage space complexities grow
exponentially with the number of cores.  Teodorescu {\em et
al.~}\cite{Teodorescu08} developed a linear programming method to find
the best DVFS settings under power constraints.  However, they assumed
power is linearly dependent on the core frequency, which is often a
poor approximation.  Meng {\em et al.~}\cite{Meng08} developed a
greedy algorithm that starts with maximum speeds for all cores and
repeatedly selects the neighboring lower global power mode with the
best $\Delta_{power}/\Delta_{perf}$ ratio.  The algorithm may traverse
the entire space of power mode combinations.  Winter {\em et
al.~}\cite{Winter10} improved this algorithm using a max-heap
data structure and reduced the complexity to $O(FN \log N)$, where $F$
is the number of core frequencies and $N$ is the number of cores.
They also developed a heuristic that runs in only $O(N \log
N)$ time.  Bergamaschi {\em et al.~}\cite{Bergamaschi08} formulated a
non-linear optimization and solved it via the interior-point method.
The method usually takes many steps to converge and its average
complexity is a high polynomial in the number of cores.
% RB: Not directly related.
% Instead of having a power budget, Ghasemazar {\em et
% al.}~\cite{Ghasemazar10} considered minimizing power under average
% throughput constraints. They also assume a simple constant
% power-frequency model.

Table~\ref{tb.compareMethods} lists some representative works and
their time complexity. We also contrast them with \powerscale as a
preview.

\begin{table}[t]  
\renewcommand{\arraystretch}{1.1}
\centering
\begin{tabular}{ccc}
\toprule
Method & Complexity & Mem. DVFS \\ \midrule
Exhaustive \cite{Isci06} & $\sim O(F^N)$ & No\\ 
Numeric Opt.\cite{Bergamaschi08, Teodorescu08} & $\sim O(N^4)$ for LP & No \\   
Heuristics \cite{Meng08, Winter10} & $\sim O(FN \log N)$ & No \\ 
\powerscale & $O(N \log M)$ & Yes \\	
	 \bottomrule
  \end{tabular} \\
\vspace{.1in}
\caption{\small Comparison of \powerscale and existing approaches. {\powerscale scales linearly
with the number of cores, while also managing memory power.}}
\label{tb.compareMethods}
\vspace{-.3in}
\end{table}

\noindent{\bf Control-theoretic approaches.}  Mishra {\em et
al.}~\cite{Mishra10} studied power management in multi-core CPUs with
voltage islands. They assumed that the power-frequency model (power
consumption as a function of the cores' frequencies) is fixed for all
islands, which may be inaccurate under changing workload dynamics.  Ma
{\em et al.}~\cite{Ma11} used a method that stabilizes the power
consumption by adjusting a frequency quota for all cores. In a similar
vein, Chen {\em et al.}~\cite{Chen11} used a control-theoretic method
along with (idle) memory power management via rank
activation/deactivation.  Unfortunately, rank activation/deactivation
is too slow for many applications \cite{Meisner11}.  Moreover,
\cite{Ma11, Chen11} require a linear power-frequency model, which may
cause under- and over-correction in the feedback control due to poor
accuracy.  This may lead to large power fluctuations, though the
long-term average power is guaranteed to be under the budget.

\noindent{\bf Other related works.} Shen {\em et al.}~\cite{Shen13} recently
considered power capping in servers running interactive applications.
They used model-based, per-request power accounting and CPU throttling
(not DVFS) for requests that exceed their fair-share power allocation
(each request is given the same power budget).  Thus, their notion of
fairness relates to the power consumption, not the performance, of
different requests. Ge {\em et al.}~\cite{Ge07} developed a runtime system on CPU
for power aware HPC computing. However they do not consider the impact on memory. Sarood {\em et al.}~\cite{Sarood14} proposed a software-based online resource management system for building power-efficient datacenter clusters. Sasaki {\em et al.}~\cite{Sasaki13} considered power capping at the thread level. They designed a run-time algorithm to distribute power budget to each application in terms of the number of cores and operating frequency. Ma {\em et al.}~\cite{Ma11_2} studied the power budgeting for multi-core CMPs together with the L2 cache. Also recently, Jha {\em et al.}~\cite{Jha15} used local Pareto front generation, followed by global utility-based power allocation to traverse the large search space of system-wide power settings. There are also works that utilize auction theory \cite{Wang14} and machine learning approaches \cite{Cochran11}.

\noindent{\bf \powerscale contributions.}  {\em There has not been any prior
work that jointly considers CPU and memory DVFS in power capping.}

Though effective in the scenarios they targeted, the prior works in
power capping are computationally expensive (e.g.,
~\cite{Isci06, Teodorescu08, Bergamaschi08}), assume potentially
inaccurate linear power models (e.g., ~\cite{Mishra10, Ma11, Chen11}),
require expensive offline profiling and model construction (e.g.,
~\cite{Cochran11, Petrica13}), or may be expensive in practice (e.g.,
~\cite{Muthukaruppan14}).

\powerscale differs from these works in many ways.  First, it 
selects active (DVFS) power modes for the cores and memory in tandem.
Second, it enforces a fair allocation of the power across the
applications running on the system {\em based on their performance},
i.e. an application may receive a larger share of the overall power
budget simply because it needs more power to match the performance
loss imposed on other applications.  Third, it leverages a
queuing-based performance model and a dynamically adjusting power
model to make frequency selection decisions with low time complexity.
Finally, our evaluation shows that \powerscale produces better
application performance and fairness than many prior approaches, even
when they are extended to use both CPU and memory DVFS.

\begin{figure}[t]
\centering
\includegraphics[width = 0.38\textwidth]{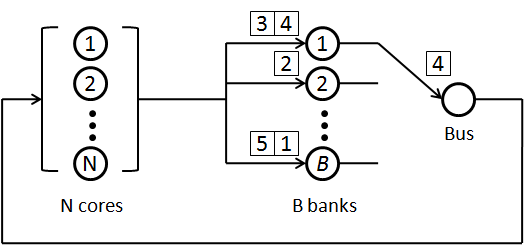}
\vspace{-.1in}
\caption{\small \powerscale's queuing model and the ``transfer
  blocking'' property.  Memory bank $1$ receives requests from cores
  $4$ and $3$. The requested data for core $4$ has been fetched and is
  being transferred on the memory bus. At the same time, bank $1$ is
  blocked from processing the request from core $3$ until the last
  request is successfully transferred to core $4$.}
\label{fig.sys_model}
\vspace{-.15in}
\end{figure}

\section{\powerscale}
\subsection{System model}
\label{sec.queuing_model}

\powerscale~models a system with $N$ (in-order) cores, $B$ memory banks,
and a common memory bus for data transfers.  (We also study
out-of-order cores in Section~\ref{sec.results}.)  Denote by
$\mathcal{N}$ the set of cores. We assume each core runs one {\em
application} and we name the collection of $N$ applications as a {\em
workload.} We use a closed-network queuing model, as depicted in
Figure~\ref{fig.sys_model}.

\vspace{.05in}
\noindent{\bf Many-core performance.}  Every core periodically issues
memory access requests (resulting from last-level cache misses and
writebacks) independently of the other cores. Though the following
description focuses on cache misses for simplicity, \powerscale~also
models writebacks as occupying their target memory banks and the
memory bus. In addition, \powerscale~assumes that writebacks happen in
the background, off the critical performance path of the cores.

\begin{figure}[t]
\centering
\includegraphics[width = 0.48\textwidth]{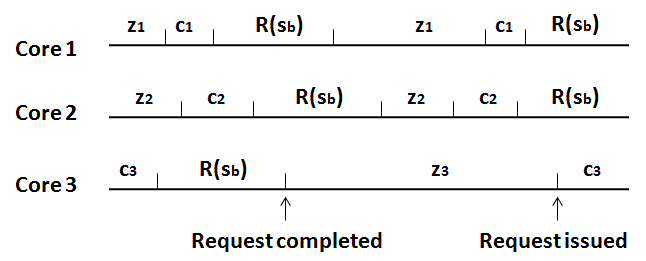}
\vspace{-.16in}
\caption{\small An example workload dynamics with $N=3$
  cores. Variables $z_i$ and $c_i$ are the think time and cache time
  for core $i$, respectively. $R(s_b)$ is the response time of the
  memory. $z_i$, $c_i$ and $R(s_b)$ are all average values. The sum
  $R(s_b) + c_i + z_i$ is the total time for one memory access of core
  $i$.}
\label{fig.samplePath}
\vspace{-.2in}
\end{figure}

After issuing a request, the core waits for the memory subsystem to
fetch and return the requested cache line before executing future
instructions.  We denote by $z_i, i \in \mathcal{N}$ the average time
core $i$ takes to generate a new request after the previous request
completes (i.e., data for the previous request is sent back to core
$i$, see Figure~\ref{fig.samplePath}). The term $z_i$ is often called
the {\em think time} in the literature on closed queuing networks
\cite{mor_book}. Further, to model core DVFS, we assume each core can
be voltage and frequency scaled independently of the other cores, as
in \cite{Kim08,Yan12}. This translates to a scaled think time: denote
by $\overline{z_i}$ the minimum think time achievable at the maximum
core frequency. Thus, the ratio $\overline{z_i}/z_i \in [0, 1]$ is the
frequency scaling factor: setting frequency to the maximum yields $z_i
= \overline{z_i}$. The minimum think time depends on the application
running on the corresponding core and may change over time. 
\powerscale takes the minimum think time $\overline{z_i}$ as
an input. Determining the frequency for core $i$ is equivalent to
determining the think time $z_i$.  We assume there are $F$
frequency levels for each core.

We assume the shared last-level cache (L2) sits in a separate voltage
domain that does not scale with core frequencies. According to our
detailed simulations, changing core frequencies does not significantly
change the per-core cache miss rate. Thus, for simplicity, we model
the average L2 {\em cache time} $c_i$ for each core $i$ as independent
of the core frequency.

\vspace{.05in}
\noindent{\bf Memory performance.}  Each of the $B$ memory banks serves
requests that arrive within its address range.
% RB: No need for this assumption.
% We assume requests arrive randomly and uniformly across all $B$
% banks. This assumption is consistent with the common practice of
% interleaving cache lines across banks.
After serving one request, the retrieved data is sent back to the
corresponding core through the common bus that is shared by all memory
banks. The bus is used in a first-come-first-serve manner: any request
that is ready to leave a bank must queue behind all other requests in
other banks that finish earlier before it can acquire the bus.
Furthermore, each memory bank cannot process the next enqueued request
until its current request is transferred to the appropriate core
(cf.~Figure~\ref{fig.sys_model}).  In queuing-theoretic terminology,
this memory subsystem exhibits a ``transfer blocking'' property
\cite{Akyildiz88, Balsamo01}.  In Figure~\ref{fig.sys_model}, we
illustrate the transfer blocking property via an example.

An important performance metric for the memory subsystem is the {\em
mean response time}, which is the average amount of time a request
spends in the memory (cf.~Figure.~\ref{fig.samplePath}). To the best
of our knowledge, no closed-form expression exists for the mean
response time in a queuing system with the transfer blocking
property. Instead of deriving an explicit form for the mean response
time, \powerscale~uses the following approximation.

When a request arrives at a bank, let $Q$ be the expected number of
requests enqueued at the bank (including the newly arrived
request). When the request has been processed and is ready to be sent
back to the requesting core, let $U$ be the expected number of
enqueued requests waiting for the bus, including the departing request
itself. Denote by $s_m$ the average memory access time at each
bank. Denote by $s_b$ the bus transfer time. \powerscale~approximates
the mean response time of the memory subsystem as:
\begin{align}
\label{eq.R}
R(s_b) \approx Q(s_m + U s_b).
\end{align}

% RB: This is obvious.
% By denoting $R(s_b)$, we emphasize that the response time is a
% function of bus transfer time.  
A previous study \cite{Deng12b} has found this equation to be a good
approximation to the response time of the memory subsystem.

The memory DVFS method is based on MemScale \cite{Deng11}, which
dynamically adjusts memory controller, bus, and dual in-line memory
module (DIMM) frequencies.  Although these memory subsystem
frequencies are adjusted together, we simplify the discussion by
focusing on adjusting only the bus frequency. This translates to a
scaled bus transfer time. Denote by $\overline{s_b}$ the minimum bus
transfer time at the maximum bus frequency -- the ratio
$\overline{s_b} / s_b \in [0, 1]$ is the bus frequency scaling
factor. We assume the bus frequency can take $M$ values. In the
\powerscale~algorithm, the minimum bus transfer time $\overline{s_b}$
is used as an input, and determining a frequency for the memory is
equivalent to determining the transfer time $s_b$.

\vspace{.05in}
\noindent{\bf Power models.}  Using our detailed simulator 
(Section~\ref{sec:evaluation}), we study the power consumption of
cores and the main memory serving different workloads.
\iffalse
In Figure~\ref{fig.CPUPwrCurve}, we plot the power consumption of a
single core when running a CPU-bound or a memory-bound workload on a
16-core system, as a function of the normalized core frequency.  As we
can see, the power profile of the core depends heavily on the
characteristics of the workload. Many prior papers e.g.~\cite{Ma11,
Teodorescu08} used simple models (e.g.,~assuming the power is always
linearly dependent on the frequency) that do not account well for
different workload characteristics.  Although these models can be
accurate for some workloads, they may be highly inaccurate for others.
Inaccurate power models can lead to control oscillation and
performance degradation, as we show in Section~\ref{sec:evaluation}.

\begin{figure}
\centering
\includegraphics[width = 0.5\textwidth]{figures/cpuPwrPlot.eps}
%\vspace{-.3in}
\caption{\small Core power when running a CPU- or a memory-bound workload.
The frequency is normalized to the maximum value.}
\label{fig.CPUPwrCurve}
\vspace{-.1in}
\end{figure}

Motivated by Figure~\ref{fig.CPUPwrCurve}
\fi
We model the power drawn by core $i$ as
\begin{align}
\label{eq.CPUPowerModel}
P_{i} \left (\frac{\overline{z_i}}{z_i} \right )^{\alpha_i} + P_{i, static},
\end{align}
where $P_{i}$ is the maximum voltage/frequency-dependent
power consumed by the core, $\alpha_i$ is some exponent typically
between 2 and 3, and $P_{i, static}$ is the static
(voltage/frequency-independent) power the core consumes at
all times.
% \yanpei{We note that
% the first term in Equation~\ref{eq.CPUPowerModel} captures the power
% that is frequency and/or voltage dependent.}
At runtime, \powerscale
periodically recomputes $P_{i}$ and $\alpha_i$ by using power
estimates for core $i$ running at different frequencies, and solving
the instances of Equation~\ref{eq.CPUPowerModel} for these
parameters. We note that many prior papers e.g.~\cite{Ma11,
Teodorescu08} used simple models (e.g.,~assuming the power is always
linearly dependent on the frequency) that do not account well for
different workload characteristics.

% RB: \powerscale~ could be much more complex for heterogeneous cores,
% e.g., what if cores had different numbers of possible frequency settings.
% (Note that for simplicity we assume cores are homogeneous, i.e., the
% $P_{core}$ is the same across all cores.  \powerscale~ can also model
% heterogeneous cores.)  
% RB: Static power is becoming increasingly significant.
% However, we find the value to be very small thus negligible.
\iffalse
In Figure~\ref{fig.MEMPwrCurve}, we plot the power consumption of the
memory in a 16-core system, when all cores run a CPU-bound or a
memory-bound workload, as a function of the normalized memory
frequency.  As we can see, the power profile of the memory also
depends heavily on the workload.

\begin{figure}
\centering
\includegraphics[width = 0.5\textwidth]{figures/memPwrPlot.eps}
%\vspace{-.3in}
\caption{\small Memory power when running a CPU- or a memory-bound workload. 
The frequency is normalized to the maximum value.}
\label{fig.MEMPwrCurve}
%\vspace{-.2in}
\end{figure}
\fi

We model the memory power as
\begin{align}
\label{eq.MemPowerModel}
P_{m} \left( \frac{\overline{s_b}}{s_b} \right)^{\beta} + P_{m, static},
\end{align}
where $P_{m}$ is the maximum memory power. In practice, we observe
that the exponent $\beta$ is close to $1$. This is because we only
scale the frequency and not the voltage of the memory bus and DIMMs.
The memory also consumes some static power $P_{m, static}$ that does
not vary with the memory frequency.  At runtime, \powerscale
periodically recomputes $P_{m}$ and $\beta$ by using power estimates
for the memory running at different frequencies, and solving the
instances of Equation~\ref{eq.MemPowerModel} for the parameters. 

We include all the sources of power consumption that do not vary
with either core or memory frequencies into a single term $P_{s}$.
This term includes the static power of all cores $\sum_i P_{i,
static}$, the memory's static power $P_{m, static}$, the memory
controller's static power, the L2 cache power, and the power consumed
by other system components, such as disks and network interfaces.

% RB: We could add this text, but I don't think that it is necessary.
% We show that our power model is accurate under dynamically changing
% workloads in our longer technical report \cite{Liu14}.
To study the accuracy of our power model under dynamically changing
workloads, we simulate both CPU- and memory-bound jobs and find that
the modeling error is less than $10\%$.

\vspace{.05in}
\noindent{\bf Model discussion.} 
By making $z_i$ represent the time
between two consecutive {\em blocking} memory accesses, \powerscale's~model can easily adapt to out-of-order cores with multiple
outstanding misses per core; assuming
non-blocking accesses are off the critical path, just as cache
writebacks. We discuss our out-of-order implementation in
Section~\ref{sec.results}. \powerscale~can also easily adapt to multiple controllers
by considering different response times for different controllers.  In
this scenario, the probability of each core using each controller
(i.e., the access pattern) has to be considered.  We defer the
discussion of multiple controllers to Section~\ref{sec.results}.

\subsection{Optimization and algorithm}
\label{sec.optimization}
\powerscale's goal is to maximize the applications' performance under
a full-system power budget.  Importantly, \powerscale seeks to fairly
allocate the budget across the cores (applications) and memory, so
that all applications degrade by the same fraction of their maximum
performance as a result of the less-than-peak power.  Thus,
\powerscale seeks to prevent ``performance outliers'', i.e. applications
that get degraded much more than others.

% Specifically, the desired features
% should include
% \begin{itemize}
% \item[1)] Being able to dynamically adjust power budget to different cores and/or memory based on their performance needs.
% \item[2)] Avoid performance outliers: no core should be running too slow to save power for (in favor of) other cores.
% \item[3)] Respect overall system power budget.
% \item[4)] Maximize the throughput as much as possible. 
% \end{itemize}
% It is important to note that some objectives can be conflicting with
% others. For instance the policy which maximizes the throughput (in
% terms of number of instructions committed by the system in a given
% period) may favor cores which are more CPU-bound. The policy may
% allocate more power to those cores, causing memory-bound cores to
% starve in power and become performance outliers.

Due to the convenience of the queuing model
(cf.~Figure~\ref{fig.samplePath}), we use the time interval between
two memory accesses (we call it {\em turn-around time}, i.e., $z_i +
c_i + R(s_b)$) as the performance metric.  Since a certain number of
instructions is executed during a given think time $z_i$, the shorter
the turn-around time is, the higher the instruction throughput and
thus the better the performance.  Based on this metric, we propose the
following optimization for \powerscale.
\begin{align}
\label{eq.obj}
\textrm{Maximize} &\quad D \\
\label{eq.perfConstraint}
\textrm{subject to} &\quad \frac{z_i + c_i + R(s_b)}{\overline{z_i} + c_i + R(\overline{s_b})} \leq 1/D \quad \forall~i \in \mathcal{N} \\
\label{eq.powerConstraint}
&\quad \sum_i P_i \left(\frac{\overline{z_i}}{z_i}\right)^{\alpha_i} + P_m \left(\frac{\overline{s_b}}{s_b}\right)^{\beta} + P_s \leq B \overline{P} \\
\label{eq.minConstraint}
&\quad s_b \geq \overline{s_b}, \quad z_i \geq \overline{z_i}, \quad s_b, z_i \in \mathbb{R} \quad \forall~i \in \mathcal{N} 
\end{align}
The optimization is over $z_i$ and $s_b$. The
objective is to maximize the performance (or to minimize the performance
degradation $1/D$ as much as possible). Constraint
\ref{eq.perfConstraint} specifies that each core's average turn-around
time can only be at most $1/D \geq 1$ of the minimum average
turn-around time for that core.  (Recall that a higher turn-around
time means lower performance.)  To guarantee fairness, we apply the
same upper-bound $1/D$ for all cores with respect to their best
possible performance (highest core and memory frequencies). 
Constraint \ref{eq.powerConstraint}
specifies that the total power consumption (core power plus memory
power plus system background power) should be no higher than the power
budget.  The budget is expressed as the peak full-system power
$\overline{P}$ multiplied by a given budget fraction $0 < B \leq 1$.
The constraints \ref{eq.minConstraint} specify the range of each
variable. Since the objective function and each constraint are convex,
the optimization problem is convex.  %Table~\ref{tb.symbols} lists all
%notations for easy reference.

\iffalse
\begin{table}[t]
\small
	\renewcommand{\arraystretch}{1.1}
	\centering
	\begin{tabular}{|l|l|}
	\hline
	{\bf Notation} & {\bf Description}\\
	\hline
	$\mathcal{N}$ & set of CPU cores \\
        $N$ &  number of CPU cores \\
	$\overline{z_i}$ & minimum average think time for core $i$ \\
	$z_i$ $(\dagger)$ & average think time for core $i$ \\
	$c_i$ & average cache time for core $i$ \\
	$P_{i}$ & maximum power for core $i$ \\
	$P_{i, static}$ & core $i$ static power \\	
	$M$ & number of memory frequency levels \\
	$F$ & number of core frequency levels \\
	$\overline{s_b}$ & minimum bus transfer time \\
	$s_b$ $(\dagger)$ & bus transfer time \\
	$s_m$ & average memory access time \\
	$P_{m}$ & maximum power for memory \\
	$P_{m, static}$ & memory static power \\
	$P_s$ & power draw that is independent of voltage/frequency \\
	$R(s_b)$ & average response time of the memory subsystem \\
	$\alpha_i$ & power-frequency relation exponent for core $i$ \\
	$\beta$ & power-frequency relation exponent for memory \\	
	$\overline{P}$ & peak full-system power \\
	$D$ & target performance degradation \\
	$B$ & power budget fraction of the full-system power\\	
	\hline
	\end{tabular} \\ 
%\vspace{-.05in}
	\caption{\small List of notations. Variables with the $(\dagger)$
          sign are the outputs of \powerscale. Others are
          inputs.}
	\label{tb.symbols}
%\vspace{-.1in}
\end{table}
\fi

Note that the optimization problem is constrained by the 
overall system budget. However, it can be extended to capture
per-processor power budgets by adding a constraint similar to
constraint \ref{eq.powerConstraint} for each processor.

\powerscale solves the optimization problem for $z_i$ and $s_b$, and
then sets each core (memory) frequency to the value that, after
normalized to the maximum frequency, is the closest to $\overline{z_i}
/ z_i$ ($\overline{s_b} / s_b$).  For the cores and memory controller,
a change in frequency may entail a change in voltage as well. 
Thus, the power consumed by each core and memory is always dynamically
adjusted based on the applications' performance needs.  The coupling
of the objective in line \ref{eq.obj} and constraint
\ref{eq.perfConstraint} seeks to minimize the performance degradation
of the application that is furthest away from its best possible
performance.  Since each core has its own minimum turn-around time and
the same upper-bound proportion is applied to all cores, we ensure
fairness among them and mitigate the performance outlier problem.

The optimization problem can be solved quickly using numerical
solvers, such as CPLEX.  However, the problem can be solved
substantially faster using the following observations.
\begin{theorem}
\label{thm.equality}
Suppose the solution $D^*$, $s_b^*$ and $z_i^*, i \in \mathcal{N}$ are
the optimal solution to the optimization problem. Then, inequalities
\ref{eq.perfConstraint} and \ref{eq.powerConstraint} must be
equalities.
\end{theorem}
% RB: I selected a different set of texts to include.  See what you
% think.  My selections allow us to avoid referencing a technical 
% report, while staying at roughly the same length.
% We present the proof in our longer technical report \cite{Liu14}.
%
\iftrue
{\em Proof.}  We first show that constraint \ref{eq.powerConstraint}
must be an equality.
% \ricardo{Though $z_i$ and $s_b$ are fine-grain, the knob to
% change them is coarse-grained and very discrete.  I don't see how one
% can guarantee that constraint \ref{eq.powerConstraint} must be an
% equality under these conditions.}  \yanpei{Equalities may not hold for
% coarse-grained/discrete frequencies, however the theorem is about
% $z_i$ and $s_b$, not about frequencies. The optimization is also set
% up as optimizing over $z_i$ and $s_b$, not directly frequencies.}
Suppose otherwise, then we can always reduce the optimal bus speed
$s_b^*$ such that the performance of each core is improved (because of
the decrease in $R(s_b^*)$).  As a result, we can achieve a better
objective, larger than $D^*$.  This leads to a contradiction.  Thus,
the power budget constraint must be an equality.

Now, we show that constraint \ref{eq.perfConstraint} must also be an
equality.  Suppose otherwise, i.e. there exists a $j$ such that
constraint \ref{eq.perfConstraint} is strictly smaller than $1/D^*$.
Then, we can increase $z_j^*$. The power budget saved from this core
can be redistributed to other cores that have equalities in constraint
\ref{eq.perfConstraint}.  As a result, we can achieve an objective 
that is larger than $D^*$.  This leads to a contradiction as well.
$\square$
\fi

Theorem~\ref{thm.equality} suggests that the optimal solution must
consume the entire power budget and each core must operate at $1/D$
times of its corresponding target.  With constraints
\ref{eq.perfConstraint} and \ref{eq.powerConstraint} as
equalities, the optimal think time $z_i$ can be solved in linear time
$O(N)$ for a given bus time $s_b$. This is because $z_i$ can be
written as
\begin{align}
\label{eq.1}
z_i = \frac{\overline{z_i} + c_i + R(\overline{s_b})}{D} - c_i - R(s_b).
\end{align}
We then substitute Equation \ref{eq.1} into constraint
\ref{eq.powerConstraint}, and solve for $D$ using the equality
condition for constraint \ref{eq.powerConstraint}.  Then, all optimal
$z_i$ can be computed in linear time using Equation \ref{eq.1}.

We can then exhaustively search through $M$ possible values for $s_b$
to find the globally optimal solution.  However, since the
optimization problem is convex, we only need to find a local optimal.
Since we can find an optimal solution for each bus transfer time
$s_b$, we can simply perform a binary search across all $M$ possible
values for $s_b$ to find the local optimal.  This results in the $O(N
\log M)$ algorithm shown in Algorithm~\ref{alg.alg1}. 

We cannot quantitatively compare \powerscale to CoScale \cite{Deng12b}, as
they solve different problems; CoScale would have to be redesigned for
power capping. However, we can qualitatively compare how efficiently
they explore the possible power mode configurations via their time
complexities. CoScale's complexity is $O(M + FN^2)$, where $F$ is the
number of core frequencies, i.e. it scales poorly with the number of
cores.

\begin{algorithm}[t]
\caption{\powerscale~$O(N \log M)$ algorithm}
\label{alg.alg1}
\begin{algorithmic}[1]
\STATE {\bf Inputs}: $\{P_i\}$, $\{\alpha_i\}$, $P_m$, $\beta$, $P_s$, $\{\overline{z_i}\}$, $\overline{s_b}$, $Q$, $U$, $s_m$, $B$, $\overline{P}$ and an ordered array of $M$ candidate values for $s_b$.
\STATE {\bf Outputs}: $\{z_i\}$ and $s_b$
\STATE Let $\ell := 0$ and $r := M-1$.
\WHILE {$\ell \neq r$}
\STATE $m := (\ell+r) / 2$.
\STATE Solve the optimal $D$ for the $m^{th}$ $s_b$ value.
\STATE Solve the optimal $D$ for the $(m \pm 1)^{th}$ $s_b$ values. Let the optimal $D$ be denoted as $D^+$ and $D^-$ respectively. 
\IF {$D < D^+$}
\STATE $\ell := m$
\ELSIF {$D^- > D$}
\STATE $r := m$
\ELSE
\STATE {\bf break} 
\ENDIF
\ENDWHILE 
\STATE Set each core (memory) frequency to the closest frequency to $\overline{z_i} / z_i$ ($\overline{s_b} / s_b$) after normalization.
\end{algorithmic}
\end{algorithm}

\subsection{Implementation}
\label{sec.implementation}
\noindent{\bf Operation.} \powerscale splits time into fixed-size
epochs of several milliseconds each. It collects performance counters
from each core $300~\mu s$ into each epoch, and uses them as inputs to
the frequency selection algorithm. We call this $300~\mu s$ the {\em
profiling phase} and find its length enough to capture the latest
application behaviors. During the profiling phase, the applications
execute normally.

Given the inputs, the OS runs the \powerscale~algorithm and may
transition to new core and/or memory voltage/frequencies for the
remainder of the epoch.  During a core's frequency transition, the
core does not execute instructions, but other cores can operate
normally.  To adjust the memory frequency, all memory accesses are
temporarily halted, and PLLs and DLLs are re-synchronized.  The core
and memory transition overheads are small (tens of microseconds), thus
negligible compared to the epoch length.
% At the end of the epoch, key parameters such as $Q$, $U$ and
% $\overline{z_i}$ are assessed again.

\vspace{.05in}
\noindent{\bf Collecting input parameters.} Several key
\powerscale~parameters, such as $P_{i}$, $\alpha_i$, $P_m$, $\beta$,
the minimum think time $\overline{z_i}$, and queue sizes $Q$ and $U$
come directly or indirectly from performance counters.  Now, we detail
how we obtain the inputs to the algorithm from the counters.  To
compute $\overline{z_i}$, we use
\begin{align}
\label{eq.toMinZi}
TPI_i \times \frac{TIC_i}{TLM_i},
\end{align}
and scale Equation \ref{eq.toMinZi} by the ratio between the maximum
frequency and the frequency used during profiling. $TPI_i$ is the {\em
Time Per Instruction} for core $i$ during profiling, $TIC_i$ is the
{\em Total Instructions Executed} during profiling, and $TLM_i$ is the
{\em Total Last-level Cache Misses} (or number of memory accesses)
during profiling. The ratio between $TIC_i$ and $TLM_i$ is the average
number of instructions executed between two memory accesses.
% RB: The following is incorrect.  An in-order processor can have
% multiple outstanding memory accesses, as long as instructions execute
% in-oder on the data brought from memory.  The reason our modeling
% works is that we only allow one outstanding miss per core.
%
% (recall that after an in-order processor issues a request to the
% memory, it cannot execute further instructions until it receives the
% data from the memory).

We obtain $P_{i}$, $\alpha_i$, $P_{m}$, and $\beta$ as described when
they were first introduced above.  \powerscale keeps data about the
last three frequencies it has seen, and periodically recomputes these
parameters.
% \ricardo{One of the
% frequencies we can use is the profiling frequency, when/how do we get
% the other?  Do we change frequency dynamically just to observe?  We
% need to explain this more carefully.} \yanpei{In the simulator we are
% able to pull out the predicted power values at many different
% frequencies from the profiler. In real implementation, I would say 1)
% the power model doesn't need to be changed frequently and 2) one can
% always keep the most recent $3$ power values at 3 different
% frequencies in a data structure (not limited to the immediate past
% profiling phase) to recompute the parameters. If the profiling phase
% use a different frequency, retire the oldest values from the data
% structure and insert this new value. If the profiling phase use one of
% the 3 existing frequencies in the data structure, update the power
% value for that frequency.}

To obtain $Q$ and $U$, we use the performance counters proposed by
\cite{Deng11}. The counters log the average queue sizes at each
memory bank and bus. 
% By the assumption that requests are uniformly distributed across 
% all $B$ memory banks, we 
We obtain $Q$ by taking the average queue size across all banks. We
obtain $U$ directly from the corresponding counter.

To obtain $s_m$, we take the average memory access time at each bank
during the profiling phase.  The minimum bus transfer time
$\overline{s_b}$ is a constant and, since each request takes a fixed
number of cycles to be transferred on the bus (the exact number
depends on the bus frequency), we simply divide the number of cycles
by the maximum memory frequency to obtain $\overline{s_b}$.

% RB: This paragraph is wrong.  McPAT is used in our simulations and
% has nothing to do with the actual implementation of FastScale.
%
% To profile the power values for $P_{core}$, we use McPAT
% \cite{Li2009}. Core static power, L2 cache power and memory controller
% power are also profiled using McPAT. Memory power $P_{mem}$ and memory
% static power are obtained using the power model described in
% \cite{Micron07}.
All background power draws (independent of core/\-memory frequencies or
workload) can be measured and/or estimated statically.
%
%To avoid repeated overhead in computing exponents $\alpha_i$ and
%$\beta$ (cf.~Table~\ref{tb.symbols}), we treat them as fixed
%constants.  As these parameters do not change significantly (they are
%mostly determined by the architecture and the type of the workload),
%we pre-compute them offline. 

%\vspace{.05in}
%\noindent{\bf Performance management.} 
%We now discuss how we set the performance constraint $T_i$.  We set $T_i$ to be the expected performance each core would achieve if the system runs at the maximum core and memory frequencies (called baseline system). Specifically, we set
%\begin{align}
%\label{eq.Ti}
%T_i := \frac{L}{\overline{z_i} + c_i + R(\overline{s_b})}.
%\end{align}
%This constraint ensures that in every epoch, each core has its own performance target respect to the {\em same core} in the baseline system. Since in \eqref{eq.perfConstraint}, all core is expected to perform $D$ times their corresponding targets, it effectively reduces the chance of having outlier cores which may always choose minimum frequency to save power for other cores. In Section~\ref{sec:evaluation}, we will demonstrate this simple performance management method provides the best per core performance among most state-of-the-art designs. 

\begin{table}[t]
\begin{center}
\begin{small}
\vspace{-.05in}
{
\renewcommand{\arraystretch}{1}
   \begin{tabular}{ | p{0.7 cm} | p{2.4 cm} | p{4 cm} | }
   \hline
   \multicolumn{2}{|l|}{Feature} & \multicolumn{1}{l|}{Value}\\
   \hline
   \multicolumn{2}{|l|}{CPU cores} & \multicolumn{1}{l|}{$N$ in-order, single thread, 4GHz}\\
   \multicolumn{2}{|l|}{} & \multicolumn{1}{l|}{Single IALU IMul FpALU}\\
   \multicolumn{2}{|l|}{} & \multicolumn{1}{l|}{FpMulDiv}\\
   \multicolumn{2}{|l|}{L1 I/D cache (per core)} & \multicolumn{1}{l|}{32KB, 4-way, 1 CPU cycle hit}\\
   \multicolumn{2}{|l|}{L2 cache (shared)} & \multicolumn{1}{l|}{16MB, $N$-way, 30 CPU cycle hit}\\
   \multicolumn{2}{|l|}{Cache block size} & \multicolumn{1}{l|}{64 bytes} \\
   \multicolumn{2}{|l|}{Memory configuration} & \multicolumn{1}{l|}{4 DDR3 channels for 16/32 cores}\\
	\multicolumn{2}{|l|}{} & \multicolumn{1}{l|}{8 DDR3 channels for $64$ cores}\\   
   \multicolumn{2}{|l|}{} & \multicolumn{1}{l|}{8 2GB ECC DIMMs}\\
   \hline
   \multirow{6}{10mm}{\!\!\! Time}        & tRCD, tRP, tCL     & 15ns, 15ns, 15ns  \\
				   & tFAW		& 20 cycles  \\
				   & tRTP		& 5 cycles   \\
				   & tRAS		& 28 cycles \\
                                   & tRRD               & 4 cycles   \\
                                   & Refresh period     & 64ms\\
   \hline
   \multirow{7}{10mm}{\!\!\!\!\! Current} 
                                    & Row buffer & 250 (read), 250 (write) mA \\
                                    & Pre-chrg & 120 mA \\
                                    & Active standby      & 67  mA \\
                                    & Active pwrdown    & 45  mA \\
                                    & Pre-chrg standby   & 70  mA \\
                                    & Pre-chrg pwrdown & 45  mA \\
                                    & Refresh             & 240 mA \\
   \hline
   \end{tabular}
}
\end{small}
\end{center}
\caption{\small Main system settings.}
\label{tab:sim}
\vspace{-.2in}
\end{table}

\begin{table}[t]
\renewcommand{\arraystretch}{1.1}
\begin{center}
\begin{small}
%\vspace{-.05in}
{
 %\begin{tabular}{ | c | c | c | c c c c | c | c | c | c c c c |}
 \begin{tabular}{ | c | c | c | m{0.8 cm} m{0.8 cm} m{0.8 cm} m{0.9 cm} |}
 \hline
 %Name & MPKI & WPKI & \multicolumn{4}{c|}{\centering Applications (xC/4 each)}&Name & MPKI & WPKI & \multicolumn{4}{c|}{\centering Applications (x4 each)}\\
 Name & MPKI & WPKI & \multicolumn{4}{c|}{\centering Applications ($\times N/4$ each)}\\
 \hline
  ILP1 &0.37  &0.06  &vortex &gcc   &sixtrack&mesa  \\
  ILP2 &0.16  &0.03  &perlbmk&crafty&gzip  &eon     \\
  ILP3 &0.27  &0.07  &sixtrack&mesa&perlbmk&crafty\\
  ILP4 &0.25  &0.04  &vortex &gcc  &gzip &eon \\  
  MID1 &1.76  &0.74  &ammp  &gap   &wupwise &vpr     \\
  MID2 &2.61  &0.89  &astar &parser &twolf  &facerec \\
  MID3 &1.00  &0.60  &apsi  &bzip2 &ammp   &gap    \\
  MID4 &2.13  &0.90  &wupwise&vpr  &astar  &parser\\
  MEM1 &18.22  &7.92  &swim  &applu &galgel  &equake \\
  MEM2 &7.75  &2.53  &art  &milc   &mgrid &fma3d    \\
  MEM3 &7.93  &2.55  &fma3d  &mgrid &galgel &equake\\
  MEM4 &15.07 &7.31  &swim   &applu &sphinx3  &lucas  \\
  MIX1 &2.93  &2.56  &applu &hmmer &gap   &gzip     \\
  MIX2 &2.55  &0.80  &milc  &gobmk &facerec&perlbmk \\
  MIX3 &2.34  &0.39  &equake &ammp  &sjeng  &crafty  \\
  MIX4 &3.62  &1.20  &swim   &ammp  &twolf  &sixtrack  \\
  \hline
  \end{tabular}
}
\end{small}
\end{center}
\caption{\small Workload descriptions.}
\label{tab:wloads}
\vspace{-.2in}
\end{table}

\iftrue
\subsection{Hardware and software costs}

% We now discuss \powerscale's implementation cost. 
\powerscale~requires
no architectural or software support beyond that in \cite{Deng12b}.
Specifically, core DVFS is widely available in commodity hardware. 
Existing
DIMMs support multiple frequencies and can switch among them by
transitioning to powerdown or self-refresh states
\cite{Association2009}, although this capability is typically not used
by current servers.  Integrated CMOS memory controllers can leverage
existing DVFS technology.  One needed change is for the memory
controller to have separate voltage and frequency control from other
processor components. In recent Intel architectures, this would
require separating shared cache and memory controller voltage control.
In terms of software, the OS must periodically invoke \powerscale~and
collect several performance counters.
\fi

\section{Evaluation}
\label{sec:evaluation}
% In this section, we evaluate \powerscale~and compare it against
% several other DVFS-based power capping approaches.

\subsection{Methodology}
%\vspace{-.1in}

% \vspace{.05in}
\noindent{\bf Simulation infrastructure.}  We adopt the infrastructure
used in \cite{Deng12b}. 
% RB: This is shown in the table.
% We simulate 4 DDR3 channels for $16$ and $32$ cores and 8 DDR3
% channels for $64$ cores, each of which populated with two registered,
% dual-ranked DIMMs with 18 DRAM chips each.
We assume per-core DVFS, with 10 equally-spaced frequencies in the
range 2.2-4.0 GHz.  We assume a voltage range matching Intel's
Sandybridge, from 0.65 V to 1.2 V, with voltage and frequency scaling
proportionally, which matches the behavior we measured on an i7
CPU. We scale memory controller frequency and voltage, but only
frequency for the memory bus and DRAM chips.  The on-chip 4-channel
memory controller has the same voltage range as the cores, and its
frequency is always double that of the memory bus.  We assume that the
bus and DRAM chips may be frequency-scaled from 800 MHz to 200 MHz,
with steps of 66 MHz.  The infrastructure simulates in detail the
aspects of cores, caches, memory controller, and memory devices that
are relevant to our study, including memory device power and timing,
and row buffer management.  Table \ref{tab:sim} lists our default
simulation settings.

We model the power for the non-CPU, non-memory components as a fixed
10 W.
% (we show the impact of varying this percentage in Section
% \ref{sec:sensitivity}).
%
Under our baseline assumptions, at maximum frequencies, the CPU
accounts for roughly 60\%, the memory subsystem 30\%, and other
components 10\% of system power.

\noindent{\bf Workloads.} We construct the workloads by combining
applications from the SPEC 2000 and SPEC 2006 suites.  We group them
into the same mixes as \cite{Deng11,Zheng2009}.  The workload classes
are: memory-intensive (MEM), compute-intensive (ILP), compute-memory
balanced (MID), and mixed (MIX, one or two applications from each
other class).
% For each workload class, we construct $4$ workloads, each consisting
% of different applications.
We run the best 100M-instruction simulation point for each application
(selected using Simpoints 3.0).  A workload terminates when its
slowest application has run 100M instructions.  Table \ref{tab:wloads}
describes the workloads and the L2 misses per kilo-ins\-truction
(MPKI) and writebacks per kilo-instruction (WPKI) for $N = 16$.  We
execute $N/4$ copies of each application to occupy all $N$ cores.

\iffalse
Table \ref{tab:wloads} describes the
workloads we use.  We construct the workloads by combining
applications from the SPEC 2000 and SPEC 2006 suites.  We use
workloads exhibiting a range of compute and memory behavior, and group
them into the same mixes as \cite{Deng11,Zheng2009}.  The workload
classes are: memory-intensive (MEM), compute-intensive (ILP),
compute-memory balanced (MID), and mixed (MIX, one or two applications
from each other class).  The rightmost column of Table
\ref{tab:wloads} lists the application composition of each workload;
we execute $N/4$ copies of each application to occupy all $N$ cores.

We run the best 100M-instruction simulation point for each application
(selected using Simpoints 3.0 \cite{Perelman2003}).  A workload
terminates when its slowest application has run 100M instructions.
Table \ref{tab:wloads} lists the L2 misses per kilo-ins\-truction
(MPKI) and writebacks per kilo-instruction (WPKI) for $N = 16$.  In
terms of the workloads' running times, the memory-intensive workloads
tend to run more slowly than the CPU-intensive ones. 
\fi

\subsection{Results}
\label{sec.results}

% In this section we present the \powerscale results. 
We first run all workloads under the maximum frequencies to observe
the peak power the system ever consumed.  We observe the peak power
$\overline{P}$ to be 60 Watts for 4 cores, 120 Watts for 16 cores, 210
Watts for 32 cores, and 375 Watts for 64 cores.  By default, we
present results for a 16-core system in which \powerscale~is called
every $5~ms$. (The $5~ms$ epoch length matches a common OS time
quantum.) We study different epoch lengths in later sections.

\begin{figure}[t]
\centering
\includegraphics[width = 0.48\textwidth]{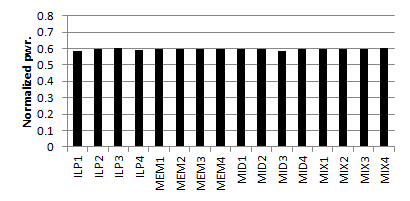}
\vspace{-.2in}
\caption{\small \powerscale~average power consumption normalized to
the peak power. Power budget is $60\%$ of the peak.}
\label{fig.paperall08pwr}
\vspace{-.1in}
\end{figure}

\noindent{\bf Power consumption.}  We first evaluate \powerscale~under
a 60\% power budget fraction, i.e. $B$ in Equation
\ref{eq.powerConstraint} equals 60\%. Figure~\ref{fig.paperall08pwr}
shows the average power spent by \powerscale~running each workload on
the 16-core system.  \powerscale~successfully maintains overall system
power just under 60\% of the peak power.

\begin{figure}[t]
\centering
\includegraphics[width=0.48\textwidth]{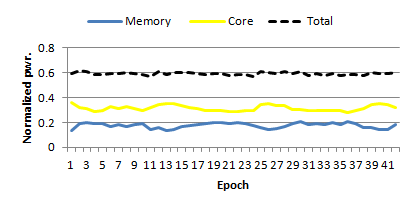}
\vspace{-.3in}
\caption{\small Normalized average power draws of cores and memory
when running MIX3 under a 60\% budget, as a function of time.}
\label{fig.Power}
\vspace{-.2in}
\end{figure}

These are overall execution averages and do not illustrate the dynamic
behavior of \powerscale.  To see an example of this behavior, in
Figure~\ref{fig.Power}, we show the breakdown of a 60\% full-system
power budget between the power consumed by the cores and by the memory
subsystem for workload MIX3, as a function of epoch number.  The
figure shows that \powerscale reacts to workload changes by quickly
repartitioning the full-system power budget.
% to different components,
% while maintaining the total power draw very near the overall
% full-system budget.

Although occasionally the average power may exceed the budget due to
workload changes, \powerscale~always maintains the power near the
budget.  As previous papers (e.g., \cite{Isci06,Lefurgy08}) have
discussed, exceeding the budget for short periods is not a problem
because the power supply infrastructure can easily handle these
violations.
% For datacenters, \cite{Wang13} suggests that short budget violations
% can be handled by drawing power from batteries.

\begin{figure}[t]
\centering
\includegraphics[width=0.48\textwidth]{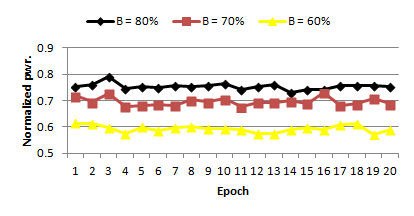}
\vspace{-.3in}
\caption{\small Normalized average power draw when running MEM3, as a
function of time and power budget.}
\label{fig.pwrDiffBudget}
\vspace{-.2in}
\end{figure}

Figure~\ref{fig.pwrDiffBudget} shows the \powerscale behavior for 3
$B$ budgets (as a fraction of the full-system peak power) for the MEM3
workload, as a function of epoch number.  The figure shows that
\powerscale corrects budget violations very quickly (within 10ms),
regardless of the budget.  Note that MEM3 exhibits per-epoch average
powers somewhat lower than the cap for $B = 80\%$.  This is because
memory-bound workloads do not consume 80\% of the peak power, even
when running at the maximum core and memory frequencies.

\begin{figure}[t]
\centering
\includegraphics[width = 0.48\textwidth]{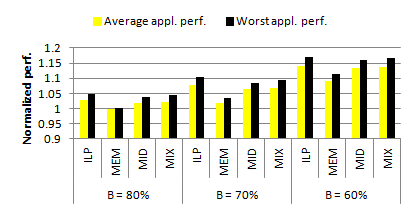}
\vspace{-.2in}
\caption{\small Average and worst application performance for each
workload class and three power budgets. }
\label{fig.perfDiffBudget}
\vspace{-.15in}
\end{figure}

\noindent{\bf Application performance.}  Recall that, under tight
power budgets, \powerscale seeks to achieve similar (percent)
performance losses compared to using maximum frequencies for all
applications.  So, where we discuss a {\em performance loss} below, we
are referring to the performance degradation (compared to the run with
maximum frequencies) due to power capping, and {\em not} to the
absolute performance.

Figure~\ref{fig.perfDiffBudget} shows the average and worst
application performance (in cycles per instruction or CPI) normalized
to the baseline system (maximum core and memory frequencies) for all
ILP, MEM, MID and MIX workloads. The higher the bar, the worse the 
performance is compared to the baseline. For each workload class, we compute
the average and worst application performance across all applications
in workloads of the class. For example, the ILP average performance is
the average CPI of all applications in ILP1, ILP2, ILP3 and ILP4,
whereas the worst performance is the highest CPI among all
applications in these workloads.  In the figure, values above 1
represent the percentage application performance loss.

This figure shows that the worst application performance differs only
slightly from the average performance.  This result shows that
\powerscale is fair in its (performance-aware) allocation of the power
budget to applications.  The figure also shows that the performance of
memory-bound workloads (MEM) tends to degrade less than that of
CPU-bound workloads (ILP) under the same power budget.  This is
because the MEM workloads usually consume less full-system power than
their ILP counterparts.  Thus, for the same power budget, the MEM
workloads require smaller frequency reductions, and thus exhibit
smaller percentage performance losses.

\begin{figure}[t]
\centering
\includegraphics[width=0.48\textwidth]{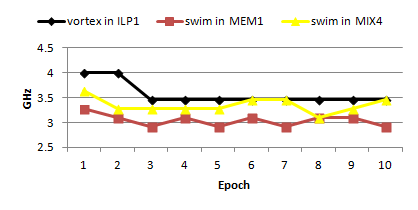}
\vspace{-.3in}
\caption{\small Core frequencies in GHz over time for cores running vortex in ILP1, swim in MEM1 and swim in MIX4. Power budget $B = 80\%$.}
\label{fig.coreFreq}
\vspace{-.2in}
\end{figure}

\begin{figure}[t]
\centering
\includegraphics[width=0.48\textwidth]{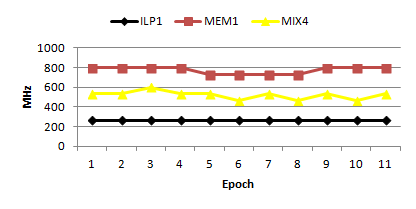}
\vspace{-.3in}
\caption{\small Memory frequencies in MHz over time when cores are running ILP1, MEM1 and MIX4. Power budget $B = 80\%$.}
\label{fig.memFreq}
\vspace{-.2in}
\end{figure}

\noindent{\bf Core/memory frequencies.} Figure~\ref{fig.coreFreq} plots the frequencies (in GHz) selected by
\powerscale~for the core running application vortex in workload ILP1,
swim in MEM1, and swim in MIX4.  Figure~\ref{fig.memFreq} plots the
memory frequencies (in MHz) selected by \powerscale~when the 16-core
system is running workloads ILP1, MEM1, and MIX4.

In the CPU-bound workload ILP1, the cores run at high frequency while
the memory runs at low frequency as expected. In the memory-bound
workload MEM1, the cores run at low frequency while the memory runs at
high frequency again as expected.  In workload MIX4, which consists of
both CPU- and memory-bound applications, memory frequencies are in the
middle of the range.  Interestingly, \powerscale selects higher core
frequencies for the core running the swim application in MIX4 than in
MEM1.  This is because in MIX4, the memory is not as busy as in MEM1,
and thus can slow down to enable higher power draws for the CPU-bound
cores.  As a result, the core running swim in MIX4 has to run faster
to compensate for the performance loss due to the slower memory
subsystem.  Since the memory power is larger than the individual core
power, sometimes it is desirable to slow down the memory and
compensate by running one or more cores faster.

\noindent{\bf \powerscale~compared with others policies.} We now compare
\powerscale~against other power capping policies.  {\em All policies
are capable of controlling the power consumption around the budget},
so we focus mostly on their performance implications.
We first compare against policies that do {\em not} use memory DVFS:
\begin{itemize}
\item {\em CPU-only.} This policy sets the core frequencies using the
\powerscale algorithm for every epoch, but keeps the memory frequency
fixed at the maximum value.  The comparison to CPU-only isolates the
impact of being able to manage memory subsystem power using DVFS.  All
prior power capping policies suffer from the lack of this capability.

\item {\em Freq-Par.} This is a control-theoretic policy from
\cite{Ma11}.  In Freq-Par, the core power is adjusted in every epoch
based on a linear feedback control loop; each core receives a
frequency allocation that is based on its power efficiency.  Freq-Par
uses a linear power-frequency model to correct the average core power
from epoch to epoch.  We again keep the memory frequency fixed at the
maximum value.
\end{itemize}

\begin{figure*}[t]
\centering
\includegraphics[width=0.95\linewidth]{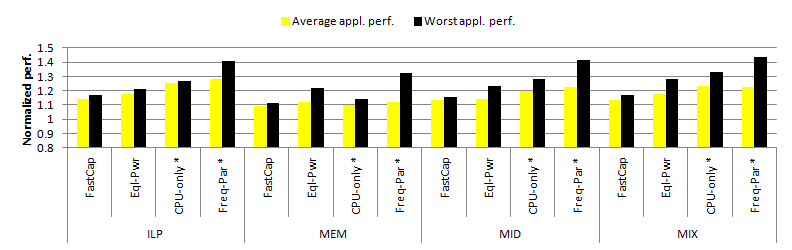}  
\vspace{-.2in}
\caption{\small \powerscale~compared with CPU-only*, Freq-Par* and
Eql-Pwr in normalized average/worst application performance. ``*''
indicates fixed memory frequency. Power budget = $60\%$.}
\label{fig.policyCompare}
\vspace{-.2in}
\end{figure*}

Figure~\ref{fig.policyCompare} shows the performance comparison
between \powerscale and these policies on a 16-core system.
\powerscale~performs at least as well as CPU-only in both average and
worst application performance, showing that the ability to manage
memory power is highly beneficial.  Setting memory frequency at the
maximum causes the cores to run slower for CPU-bound applications, in
order to respect the power budget.  This leads to severe performance
degradation in some cases.  For the MEM workloads, \powerscale and
CPU-only perform almost the same, as the memory subsystem can often be
at its maximum frequency in \powerscale to minimize performance loss
within the power budget.  Still, it is often beneficial to change the
power balance between cores and memory, as workloads change phases.
\powerscale is the only policy that has the ability to do so.

The comparison against Freq-Par is more striking.  \powerscale (and
CPU-only) performs substantially better than Freq-Par in both average
and worst application performance.  In fact, Freq-Par shows
significant gaps between these types of performance, showing that it
does not allocate power fairly across applications (inefficient cores
receive less of the overall power budget).  Moreover, Freq-Par's
linear power-frequency model can be inaccurate and causes the feedback
control to over-correct and under-correct often.  This leads to severe
power oscillation, although the long-term average is guaranteed by the
control stability.  
% RB: Removed to save space.
% Figure~\ref{fig.compareFreqParPwrTrace} shows this effect by plotting
% the average power per epoch for \powerscale~and Freq-Par running
% MIX3. Although Freq-Par maintains the average power near the budget,
% it fluctuates and repeatedly overshoots the budget.
For example, the power oscillates between 53\% and 65\% under Freq-Par
for MIX3.

Next, we study policies that use DVFS for {\em both} cores and the
memory subsystem.  These policies are inspired by prior works, but we
add \powerscale's ability to manage memory power to them:
\begin{itemize}
\item {\em Eql-Pwr.} This policy assigns an equal share of the overall
power budget to all cores, as proposed in \cite{Sharkey07}.  We
implement it as a variant of \powerscale: for each memory frequency,
we compute the power share for each core by subtracting the memory
power (and the background power) from the full-system power budget and
dividing the result by $N$.  Then, we set each core's frequency as
high as possible without violating the per-core budget.  For each
epoch, we search through all $M$ memory frequencies, and use the
solution that yields the best $D$ in Equation \ref{eq.obj}.

\item {\em Eql-Freq.} This policy assigns the same frequency to all
cores, as proposed in \cite{Herbert07}.  Again, we implement it as a
variant of \powerscale: for each epoch, we search through all $M$
and $F$ frequencies to determine the pair that
yields the highest $D$ in Equation \ref{eq.obj}.

\item {\em MaxBIPS.}  This policy was proposed in \cite{Isci06}. Its
goal is to maximize the total number of executed instructions in each
epoch, i.e. to maximize the throughput.  To solve the optimization,
\cite{Isci06} exhaustively searches through all core frequency
settings.  We implement this search to evaluate all possible
combinations of {\em core and memory} frequencies within the power
budget.  
% The system uses the combination that yields the highest
% expected throughput in the upcoming epoch.
\end{itemize}

Eql-Pwr ignores the heterogeneity in the applications' power profiles.  By splitting the core power
budget equally, some applications receive too much budget and even
running at the maximum frequency cannot fully consume it.  Meanwhile,
some power-hungry applications do not receive enough budget thus
result in performance loss.  This is most obvious in workloads with a
mixture of CPU-bound and memory-bound applications (e.g., MIX4).  As a
result, we observe in Figure~\ref{fig.policyCompare} that Eql-Pwr's
worst application performance loss is often much higher than
\powerscale's.

\begin{figure}[t]
\vspace{-.1in}
\centering
\includegraphics[width=0.48\textwidth]{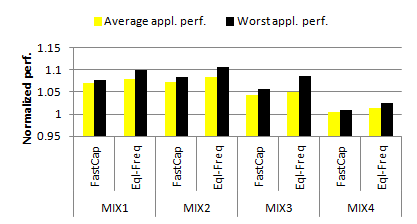}
\vspace{-.2in}
\caption{\small Normalized \powerscale~and Eql-Freq average and worst
application performance for MIX workloads on a 64-core system.  Budget
= 60\%.}
\label{fig.compareEqlFreq}
\vspace{-.2in}
\end{figure}

Eql-Freq also ignores application heterogeneity.  In Eql-Freq, having
all core frequencies locked together means that some applications may
be forced to run slowly, because raising all frequencies to the next
level may violate the power budget.  This is a more serious problem
when the workload consists of a mixture of CPU- and memory-bound
applications on a large number of cores.  To see this,
Figure~\ref{fig.compareEqlFreq} plots the normalized average and worst
application performance for \powerscale~and Eql-Freq, when running the
MIX workloads on a 64-core system.  The figure shows that Eql-Freq is
more conservative than \powerscale~and often cannot fully harvest the
power budget to improve performance.

% \begin{figure}[t]
% \centering
% \includegraphics[width=0.48\textwidth]{figures/compareFreqParPwrTrace.png}
% \vspace{-.3in}
% \caption{\small Normalized average power in \powerscale~and Freq-Par
% while running MIX3.  Power budget = $60\%$.}
% \label{fig.compareFreqParPwrTrace}
% \vspace{-.2in}
% \end{figure}

\begin{figure}[t]
\centering
\includegraphics[width=0.48\textwidth]{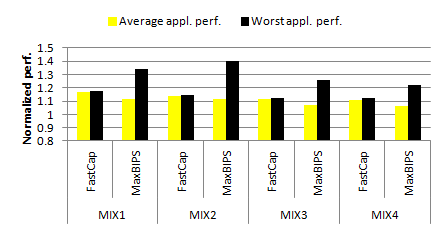}
\vspace{-.2in}
\caption{\small Normalized \powerscale~and MaxBIPS average and worst
application performance for MIX workloads on a 4-core system.  Budget
= 60\%.}
\label{fig.maxBIPScompare}
\vspace{-.2in}
\end{figure}

Finally, besides its use of exhaustive search, the main problem of
MaxBIPS is that it completely disregards fairness across applications.
Figure~\ref{fig.maxBIPScompare} compares the normalized average and
worst application performance for the MIX workloads under a 60\%
budget.  Because of the high overhead of MaxBIPS, the figure shows
results for only 4-core systems.  The figure shows that \powerscale~is
slightly inferior in average application performance, as MaxBIPS
always seeks the highest possible instruction throughput.  However,
\powerscale achieves significantly better worst application
performance and fairness.  To maximize the overall throughput, MaxBIPS
may favor applications that are more power-efficient, i.e.~have higher
throughput at a low power cost.  This reduces the power allocated to
other applications and the outlier problem occurs.  This is
particularly true for workloads that consists of a mixture of CPU and
memory-bound applications.  
% In fact, the core running memory-bound application milc in MIX2
% consistently receives low frequencies compared to other cores, making
% it a performance outlier.  
% \powerscale~demonstrates a better trade-off between maximizing the
% total throughput and limiting the outlier problem.
 
% We note that power allocation based on the ``power efficiency'' among
% cores has been extensively used in many literature. Unfortunately its
% outlier problem is often unnoticed. We thus suggest \powerscale~as an
% alternative which can effectively reduce the performance outliers.

\begin{figure}[t]
\centering
\includegraphics[width=0.48\textwidth]{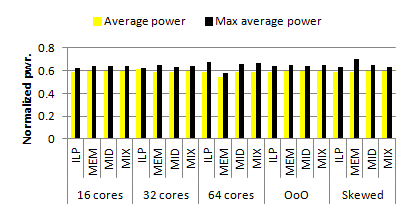}
\vspace{-.2in}
\caption{\small Normalized \powerscale average power and maximum
average power in many configurations. Power budget = 60\%.}
\label{fig.budget}
\vspace{-.2in}
\end{figure}

\noindent{\bf Impact of number of cores.} Figure~\ref{fig.budget} depicts pairs of bars for each workload class
on systems with 16, 32, and 64 cores, under a 60\% power budget.  The
bar on the right of each pair is the maximum average power of any
epoch of any application of the same class normalized to the peak
power, whereas the bar on the left is the normalized average power
{\em for the workload with the maximum average power.}  Comparing
these bars determines whether \powerscale is capable of respecting the
budget even when there are a few epochs with slightly higher average
power.  The figure clearly shows that \powerscale is able to do so
(all average power bars are at or slightly below 60\%), even though
increasing the number of cores does increase the maximum average power
slightly.  This effect is noticeable for workloads that have CPU-bound
applications on 64 cores.  In addition, note that the MEM workloads do
not reach the maximum budget on 64 cores, as these workloads do not
consume the power budget on this large system even when they run at
maximum frequencies.

\begin{figure}[t]
\centering
\includegraphics[width=0.48\textwidth]{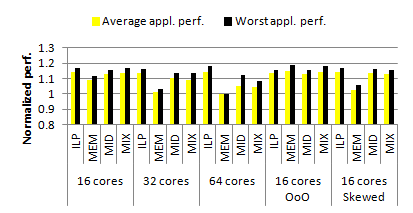}
\vspace{-.2in}
\caption{\small Normalized \powerscale average and worst application
performance in many configurations. Power budget = 60\%.}
\label{fig.performance}
\vspace{-.2in}
\end{figure}

Figure~\ref{fig.performance} also shows pairs of bars for each
workload class under the same assumptions.  This time, the bar on the
right of each pair is the normalized worst performance among all
applications in a class, and the bar on the left is the normalized
average performance of all applications in the class.  The figure
shows that \powerscale is very successful at allocating power fairly
across applications, regardless of the number of cores; the worst
application performance is always only slightly worse than the average
performance.  
%The figure also shows that application performance
%losses decrease as we increase the number of cores, especially for MEM
%workloads.  (Recall that we are comparing performance {\em losses} due
%to power capping; absolute performance cannot be compared because the
%baselines are different.)  The reason is that MEM workloads on the
%larger numbers of cores are bottlenecked by the memory subsystem even
%when they execute at maximum frequencies.

\noindent{\bf Epoch length and algorithm overhead.} By default,
\powerscale~runs at the end of every OS time quantum ($5~ms$ in our
experiments so far).  The overhead of \powerscale scales linearly with
the number of cores.  Specifically, we run the \powerscale~algorithm
for $100k$ times and collect the average time of each execution.  The
average time is $33.5~\mu s$ for $16$ cores, $64.9~\mu s$ for $32$
cores and $133.5~\mu s$ for $64$ cores. For a $5~ms$ epoch length,
these overheads are 0.7\%, 1.3\%, and $2.7\%$ of the epoch lengths,
respectively.  If these levels of overhead are unacceptable,
\powerscale can execute at a coarser granularity.  Using our
simulator, we studied epoch lengths of $10~ms$ and $20~ms$.  We find
that these epoch lengths do not affect \powerscale's ability to
control average power and performance for the applications and
workloads we consider.

\noindent{\bf Out-of-order (OoO) execution.}  Our results so far have
assumed in-order cores and \powerscale~can be easily extended
to handle the OoO executions. In \powerscale's terminology, the think time thus
becomes the interval between two core stalls (not between two main
memory accesses).  The workload becomes more CPU-bound.

We simulate
idealized OoO executions by assuming a large instruction window (128
entries) and disregarding instruction dependencies within the window.
This models an upper-bound on the memory-level parallelism (and has
no impact on instruction-level parallelism, since we still simulate a
single-issue pipeline).  
% We modify the \powerscale~models to consider the think time as the
% interval between two core stalls.
% Figure~\ref{fig.OoOPwr} shows the power consumption of running
% different workloads in our OoO mode under the $60\%$ power budget
% fraction. \powerscale~maintains the power budget throughout the
% execution.

%\begin{figure}[t]
%\includegraphics[width = 0.5\textwidth]{figures/OoOPwr.png}
%\caption{\powerscale~average power running different workloads in OoO execution normalized to the baseline. Power %budget fraction $60\%$.}
%\label{fig.OoOPwr}
%\end{figure}

% \begin{figure}[t]
% \begin{minipage}[b]{0.22\textwidth}
% \includegraphics[width = 1.0\textwidth]{figures/OoOCPI.png}
% \caption{\powerscale~average CPI comparison in OoO and in-order executions. Power budget fraction $60\%$.}
% \label{fig.OoOCPI}
% \end{minipage}
% \quad
% \begin{minipage}[b]{0.22\textwidth}
% \includegraphics[width = 1.0\textwidth]{figures/multiMCs.png}
% \caption{\powerscale~average CPI in uniform and skewed memory interleaving schemes. Power budget fraction $60\%$.}
% \label{fig.multiMC}
% \end{minipage}
% \end{figure}

Figure~\ref{fig.budget} shows four pairs of bars for the OoO
executions of the workload classes on 16 cores and under a 60\% power
budget.  The results can be compared to the bars for 16 cores on the
left side of the figure.  This comparison shows that \powerscale is
equally successful at limiting the power draw to the budget,
regardless of the processor execution mode.

Similarly, Figure~\ref{fig.performance} shows four pairs of bars for
OoO executions on 16 cores, under a 60\% budget.  These performance
loss results can also be compared to those for 16 cores on the left of
this figure.  The comparison shows that workloads with memory-bound
applications tend to exhibit higher performance losses in OoO
execution mode.  The reason is that the performance of these
applications improves significantly at maximum frequencies, as a
result of OoO; both cores and memory become more highly utilized.
When \powerscale imposes a lower-than-peak budget, frequencies must be
reduced and performance suffers more significantly.  Directly
comparing frequencies across the execution modes, we find that
memory-bound workloads tend to exhibit higher core frequencies and
lower memory frequency under OoO than under in-order execution.  This
result is not surprising since the memory can become slower in OoO
without affecting performance because of the large instruction window.
Most importantly, \powerscale is still able to provide fairness in
power allocation in OoO, as the performance losses are roughly evenly
distributed across all applications.

% we compare the average CPI of \powerscale~running in the OoO mode and
% the ones in the in-order mode on the 16-core system. We compare the
% average CPI values rather than average/worst application performance
% as OoO and in-order have different baselines.  We observe that for
% CPU-bound workloads, the average OoO CPIs are almost the same to their
% in-order counterparts.  This is because memory accesses rarely occur
% in CPU-bound workloads thus do not often fall into the same
% 128-instruction window -- the core will almost always stall at a
% memory access like in the in-order mode. In contrast, for memory-bound
% workloads, the cores stall less in OoO thus the average CPI is better
% (smaller).

% Since OoO execution makes the memory-bound workloads more CPU-bound,
% we observe that \powerscale~selects lower memory frequencies and
% higher core frequencies for the same workloads compared to the
% in-order counterparts. For instance in MEM1, the memory frequencies in
% OoO are around $533$ MHz compared to $800$ MHz in in-order
% (cf.~Figure~\ref{fig.memFreq}). The core running the swim application
% uses frequencies around $3.28$ GHz in OoO compared to $3.00$ GHz in
% in-order (cf.~Figure~\ref{fig.coreFreq}).

% Over the course of the simulation, \powerscale~always maintains the
% power budget near $60\%$ of the peak power.

\noindent{\bf Multiple memory controllers.} For \powerscale~to support multiple memory controllers (operating at
the same frequency), we use the existing performance counters to keep
track of the average queue sizes $Q$ and $U$ of each memory
controller.  Thus, different memory controllers can have different
response times (cf.~Equation \ref{eq.R}).  We also keep track of the
probability of each core's requests going through each memory
controller.  In this approach, the response time $R$ in Equation
\ref{eq.perfConstraint} becomes a weighted average across all memory
controllers and different cores experience different response times.

To study the impact of multiple memory controllers, we simulate four
controllers in our 16-core system.  In addition, we simulate two
memory interleaving schemes: one in which the memory accesses are
uniformly distributed across memory controllers, and one in which the
distribution is highly skewed.  
%In the uniformly distributed case, all
%memory controllers have roughly the same response time and all cores
%see the same response time.  In the skewed distribution, some memory
%controllers are overloaded.  

Figure~\ref{fig.budget} shows four pairs of bars for the skewed
distribution on 16 cores, under a 60\% budget.  Compare these results
to the 16-core data on the left side of the figure.  The skewed
distribution causes higher maximum power in the MEM workloads.  Still,
\powerscale is able to keep the average performance for the workload
with this maximum power slightly below the 60\% budget.

Again, Figure~\ref{fig.performance} shows four pairs of the skewed
distribution on 16 cores, under a 60\% budget.  We can compare these
performance losses to the 16-core data on the left of the figure.  The
comparison shows that \powerscale provides fair application
performance even under multiple controllers with highly skewed access
distributions.

\section{Conclusion}

In this paper we presented \powerscale, an optimization framework and
algorithm for system-wide power capping, using both CPU and memory
DVFS, while promoting fairness across applications.  The \powerscale
algorithm solves the optimization online and its complexity is the lowest among
some of the state-of-the-arts. Our
evaluation showed that \powerscale caps power draw effectively, while
producing better application performance and fairness than many
sophisticated CPU power capping methods, even after they are extended
to use of memory DVFS as well.

\section*{Acknowledgements}
We would like to thank the anonymous reviewers for suggestions that
helped improve the paper.  This work was funded in part by NSF grant
CCF-1319755. The work of Yanpei Liu was partially supported by a visitor grant from DIMACS, funded by the National Science Foundation under grant numbers CCF-1144502 and CNS-0721113.

\bibliographystyle{ieeetr}
\bibliography{paper}

\end{document}